\documentclass[12pt]{iopart}
\expandafter\let\csname equation*\endcsname\relax
\expandafter\let\csname endequation*\endcsname\relax
\usepackage{physics}
\usepackage{graphicx}

\usepackage{color}

\begin{document}

\title[Electron \textit{g}-factor of strained Ge]{Electron \textit{g}-factor of strained Ge caused by the SiGe substrate and its dependence on growth  directions}

\author{K Imakire$^1$, A Oiwa$^2$, and Y Tokura$^1$}

\address{$^1$Faculty of Pure and Applied Sciences, University of Tsukuba, Tsukuba, Ibaraki 305-8571, Japan}
\address{$^2$SANKEN, Osaka University, Ibaraki, Osaka 567-0047, Japan}
\ead{tokura.yasuhiro.ft@u.tsukuba.ac.jp}
\vspace{10pt}
\begin{indented}
\item[]October
\end{indented}

\begin{abstract}
For photon-spin conversion, the Ge hole system in a strained GeSi/Ge quantum well with a diamond structure has attracted significant attention because of the potential for a high-performance spin qubit and optical transitions ranging in telecom bands. We calculated the electron $g$-factor for strained Ge, analyzing its dependence on both the growth directions ([100], [110], and [111]) and the Ge content of the SiGe substrate using an 8-band model. Our results indicate that the absolute values of the electron $g$-factor decrease with decreasing Ge content, ranging from approximately $-3.0$ to $-1.4$ for all growth directions.
\end{abstract}

\section{Introduction}
One of the technical challenges in the optical communication of quantum information is the realization of quantum repeaters, which enable the transmission of quantum via photons over long distances using entanglement and Bell measurements. Scalable quantum repeater designs require solid-state quantum memories capable of transducing photon-based quantum information to spin-based systems \cite{Oiwa}. Vrijen and Yablonovitch had proposed spin-photon conversion using an electron system in GaAs \cite{VRIJEN2001569}, where photon information is stored in an excited electron state with nearly degenerate Zeeman sub-levels under an in-plane magnetic field. However, GaAs have nuclear spins, which complicates its application in spin qubits.

In contrast, Ge offers a better spin environment with reduced nuclear spin effects, especially when isotope engineering is applied. Ge-based spin systems have been widely studied \cite{Gequantuminfo}, including nanowires \cite{nanowire}, hut wires \cite{Watzinger2016}, and quantum wells \cite{Hendrickx2021, Lawrie}. Notably, physical properties of nanowires strongly depend on growth direction.  Additionally, there is growing interest in hexagonal crystal structures as new quantum systems \cite{Peeters2024}. The Ge hole system confined in a SiGe/Ge quantum well is a promising candidate for quantum memory due to its lighter effective mass. Coherent manipulation of hole spins has been demonstrated using spin-orbit interaction and $g$-tensor modulation. However, for efficient transduction of photon polarization information to hole spins, it is critical that the electron Zeeman splitting at the $\Gamma$ point in the conduction band is significantly larger than that of the hole in the valence band, contrary to the proposal by Vrijen and Yablonovitch \cite{VRIJEN2001569}. This energy splitting is determined by the $g$-factor, and in the case of a Ge quantum well grown on a Si$_{0.2}$Ge$_{0.8}$ substrate, the absolute value of the hole $g$-factor for an in-plane magnetic field has been reported to be between $0.16$ and $0.26$ \cite{Hendrickx2021}. The bulk electron $g$-factor at the $\Gamma$ point is theoretically predicted to be approximately $-2.6$ \cite{Roth} and experimentally observed to be $-3.0\pm0.2$ \cite{PhysRevB.2.446}. However, the strain effect in a Ge quantum well grown on a SiGe substrate is significant and the $g$-factor of electrons in strained Ge quantum wells has yet to be reported.

In this paper, we theoretically investigate the electron $g$-factor in a strained Ge layer within a SiGe heterostructure subjected to an in-plane magnetic field. We assume the Ge layer is grown on a diamond-structured SiGe substrate, where lattice mismatch induces strain. This strain effect is modeled using deformation potentials \cite{deformation}. We also explore the influence of the crystal growth direction, examining whether the $g$-factor's dependence on growth direction resembles that observed in nanowire structures. Our $g$-factor calculations exclude distant band effects and employ an 8-band model that includes the lowest conduction band, heavy hole (HH), light hole (LH), and spin split-off (SP) bands. Notably, we do not account for the quantum well confinement effect in these or the following calculations.

First, in Sec. 2, we explain the model of the strained Ge layer. In Sec. 3, we introduced the strain Hamiltonian and present the eigenenergies of the system. Section. 4 provides the $g$-factor calculations and Sec. 5 contains the discussions and conclusion.

\section{Model}
\begin{figure}[ht]
    \centering
    \includegraphics[width=0.9\linewidth]{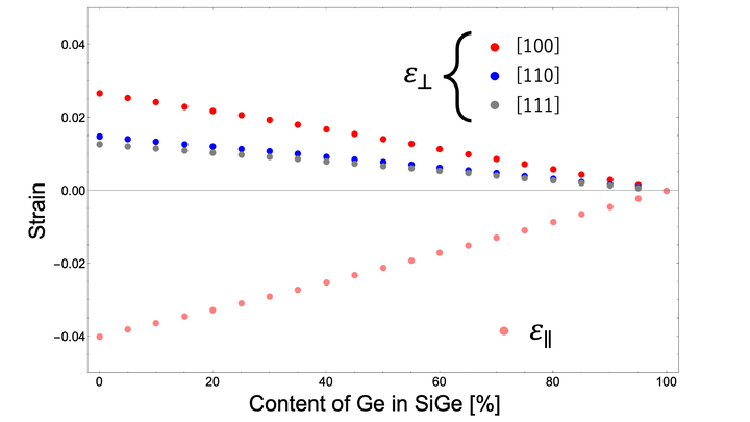}
    \caption{The Strain corresponding are shown with respect to the growth direction and the Ge content in the SiGe substrate. The calculations are performed assuming a diamond crystal structure.}
    \label{strain}
\end{figure}

We consider a Ge layer epitaxially grown on Si$_{1-x}$Ge$_x$ alloy. In this case, the Ge lattice is deformed due to the lattice mismatch between Ge and Si$_{1-x}$Ge$_x$ alloy. The lattice constant of Si$_{1-x}$Ge$_x$, $a_{\mathrm{SiGe}}$, is smaller than that of bulk Ge $a_{\mathrm{Ge}}$,\cite{GeSilattice} resulting in compressive strain on the Ge layer. The deformed position $\textbf{x}'$ after lattice mismatch can be expressed as a function of the initial position $\textbf{x}$ before deformation as follows
\begin{equation}
    \mqty(
        x'\\
        y'\\
        z'
    )=\mqty(
        x\\
        y\\
        z
    )+
    \mqty(
        \epsilon_{xx} & \epsilon_{xy} & \epsilon_{xz}\\
        \epsilon_{yx} & \epsilon_{yy} & \epsilon_{yz}\\
        \epsilon_{zx} & \epsilon_{zy} & \epsilon_{zz}
    )
    \mqty(
        x\\
        y\\
        z
    ),
\end{equation}
where $\epsilon_{ij}\;\;(i,j=x,y,z)$ represent the strain tensor elements which are symmetric ($\epsilon_{ij}=\epsilon_{ji}$). For a diamond lattice structure, the strain tensor $\epsilon$ can be written as follows,
\begin{equation}
    \epsilon=\mqty(
        \epsilon_{xx} & \epsilon_{xy} & \epsilon_{xz}\\
        \epsilon_{yx} & \epsilon_{yy} & \epsilon_{yz}\\
        \epsilon_{zx} & \epsilon_{zy} & \epsilon_{zz}
    )
    =\mqty(
        \epsilon_{\bot} & 0 & 0\\
        0 & \epsilon_{\parallel} & 0\\
        0 & 0 & \epsilon_{\parallel}
    ).
\end{equation}
We align the $z$-axis along one of the directions in the plane where the magnetic field is applied and the $x$-axis to be in the direction of crystal growth. Using linear elasticity theory, we can  relate the strain tensor $\epsilon$ to the stress tensor $\sigma$
\begin{equation}
    \sigma_{ij}=\sum_{k,l=x,y,z}C_{ijkl}\epsilon_{kl},
\end{equation}
where $C_{ijkl}$ are the elastic stiffness coefficients. From this relationship, we derive the following relations
\begin{eqnarray}
    \hspace{4cm}\epsilon_{\parallel}&=&\frac{a_{\mathrm{Ge}}-a_{\mathrm{SiGe}}}{a_{\mathrm{Ge}}}\label{epsilonpara}\\
    \hspace{4cm}\epsilon_{\bot}&=&-D\epsilon_{\parallel},\label{epsilonbot}
\end{eqnarray}
where $D$ is a constant that depends on the growth directions [100], [110] and [111]. For each direction, $D$ is
\begin{eqnarray}
    \hspace{3.5cm}D_{100}&=&\frac{2C_{12}}{C_{11}},\label{D100}\\
    \hspace{3.5cm}D_{110}&=&\frac{C_{11}+3C_{12}-2C_{44}}{C_{11}+C_{12}+2C_{44}},\label{D110}\\
    \hspace{3.5cm}D_{111}&=&\frac{11C_{11}+25C_{12}-22C_{44}}{7C_{11}+11C_{12}+22C_{44}},\label{D111}
\end{eqnarray}
where $C_{11}=C_{xxxx}$, $C_{12}=C_{xxyy}$ and $C_{44}=C_{xyxy}$ are the elastic stiffness coefficients listed in Table \ref{para}. 
For the calculation in Eq.~\eqref{epsilonpara}, the lattice parameters $a_{\mathrm{SiGe}}$ and $a_{\mathrm{Ge}}$ are taken from the data of Dismukes \etal.\cite{GeSilattice} In this context, The out-of-plane strain can be calculated by substituting Eq.~\eqref{D100},\eqref{D110},\eqref{D111}, and \eqref{epsilonpara} into Eq.~\eqref{epsilonbot}. The results are shown in Fig.~\ref{strain}. As illustrated in Fig.~\ref{strain}, the out-plane strain in the [100] direction is larger than that in the other two directions. A larger out-of-plane strain results in a smaller total volume change in the primitive cell, as it partially offsets the in-plane strain.

\section{Hamiltonian}
Under the strain induced by the Si$_{1-x}$Ge$_x$ alloy, the Hamiltonian can be expressed as
\begin{equation}
    \hat{H}_0=\frac{\hat{\textbf{p}}^2}{2m}+V(\textbf{x})+\frac{\hbar}{4m^2c^2}\qty(\hat{\sigma}\times\nabla V(\textbf{x}))\cdot\hat{\textbf{p}}+\sum_{i,j=x,y,z}\epsilon_{ij}\qty(-\frac{\hat{p}_i\hat{p}_j}{m}+U_{ij}(\textbf{x})),
    \label{hamil}
\end{equation}
where $\hat{\textbf{p}}$ is the momentum operator, $m$ is the free electron mass, $\hbar$ is the Dirac constant, $c$ is the speed of light and $\hat{\sigma}$ represents the Pauli's matrices. The periodic potential $V(\textbf{x})$ satisfies the following relation by using lattice vectors $\textbf{a}'_{\mathrm{Ge},i}(i=1,2,3)$ in the strained systems.
 \begin{equation}
 V(\textbf{x}'+\textbf{a}'_{\mathrm{Ge},i})=V(\textbf{x}')
 \end{equation}
$U_{ij}$ is defined as $U_{ij}=x_j\qty[\partial V(\textbf{x}')/\partial x'_i]_{\textbf{x}'=\textbf{x}}$.

\begin{figure}[ht]
    \centering
    \includegraphics[width=0.9\linewidth]{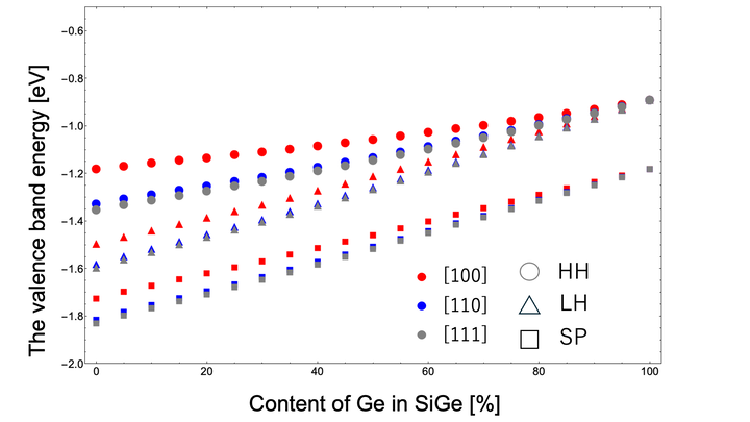}
    \caption{The eigenenergies corresponding to the heavy hole (HH), light hole (LH), and spin split-off band (SP) are shown with respect to the growth direction and the Ge content in the SiGe substrate. The calculations are performed using an 8-band model, assuming a diamond crystal structure. The origin of energy is set at the conduction band energy.}
    \label{ene}
\end{figure}

The first through third terms represent the well-known Hamiltonian of unstrained bulk Ge. The eigenvalues of unstrained Hamiltonian at the $\Gamma$ point are given as $E_c$, the conduction band energy; $E_v$, the top of valence band energy; and $E_v-\Delta$, the spin split off energy. Corresponding eigenfunctions are $\ket{S\uparrow}$,
$\ket{S\downarrow}$,$\ket{\frac{3}{2},\pm\frac{3}{2}}$,$\ket{\frac{3}{2},\pm\frac{1}{2}}$,$\ket{\frac{1}{2},\pm\frac{1}{2}}$ \cite{Cardona}. Matrix form of Eq.~\eqref{hamil}, evaluated with these eigenvalues and eigenfunctions is \cite{1130282273100471168}
\begin{equation}
    \mqty(
        0 & 0 & 0 & 0 & 0 & 0 & 0 & 0\\
        0 & 0 & 0 & 0 & 0 & 0 & 0 & 0 \\
        0 & 0 & -E'+q & 0 & {r} & 0 & 0 & {\sqrt{2}r} \\
        0 & 0 & 0 & -E'-q & 0 & {-r} & {\sqrt{2}q} & 0  \\
        0 & 0 & {r} & 0 & -E'-q & 0 & 0& {\sqrt{2}q}  \\
        0 & 0 & 0 & {-r} & 0 & -E'+q & {-\sqrt{2}r} & 0  \\
        0 & 0 & 0 &{\sqrt{2}q} & 0 & {-\sqrt{2}r} & -E'-\Delta& 0  \\
        0 & 0 & {\sqrt{2}r} & 0 & {\sqrt{2}q} & 0 & 0 & -E'-\Delta  \\
    )
    \label{sther},
\end{equation}
where $E'=E_c-E_v+(a_c-a_v)\Tr\epsilon$, $q=b(\epsilon_{\parallel}-\epsilon_{\bot})/2$, and $r=\sqrt{3}b(\epsilon_{\bot}-\epsilon_{\parallel})/2$. $a_c,a_v$ and $b$ are deformation potentials. The origin of energy is set at the conduction band energy. The eigenenergies of Eq.~\eqref{sther} are plotted in Fig.~\ref{ene}.
In Fig.~\ref{ene}, a gradual decrease can be observed in the [100] direction compared with the [110] and [111] directions. This is because a smaller volume change results in a smaller energy shift.

\section{\textit{g}-factors}
To calculate effective $g$-factor, $g^*$, we use the formula introduced by Roth \etal \cite{Roth}
\begin{equation}
    \frac{g^*}{g_0}=1-\frac{1}{mi}\sum_{n\neq S}\frac{p_{S n}^x p_{nS}^y-p_{S n}^y p_{nS}^x}{E_{n}},
    \label{geq}
\end{equation}
where $g_0$ is the free electron $g$-factor, $p_{kl}^i=\mel**{k}{\hat{p}_i}{l}$, $E_n$ is the eigenenergy of the $n$-th band measured from the conduction band energy, and the sum over $n$ excludes the conduction bands states labeled $S$. 

We choose a weak magnetic field $\mathcal{B}$ such that it does not induce interband transitions at the $\Gamma$ point. Under this condition, we can neglect the product of $\mathcal{B}$ and $\epsilon$ as a second-order infinitesimal, meaning it does not affect the linear $\mathcal{B}$ dependent term in the Hamiltonian \cite{1130282273100471168}\cite{luttinger1955motion}. Thus, we can apply Eq.~\eqref{geq} even in the strained Ge layer system, although it was originally derived for the unstrained systems.

\begin{figure}[ht]
    \centering
    \includegraphics[width=0.8\linewidth]{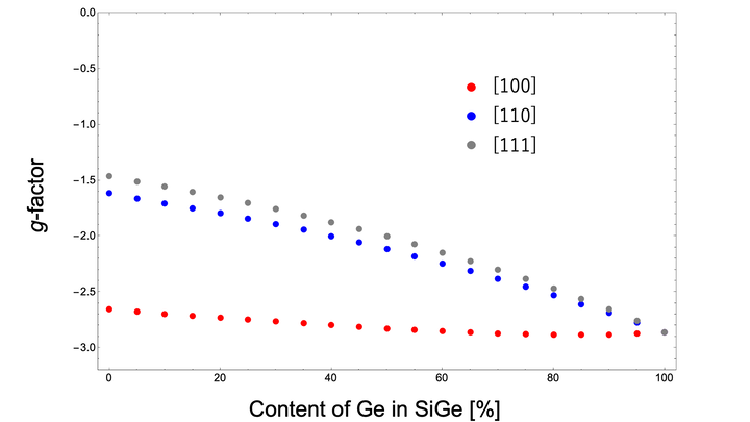}
    \caption{The electron $g$-factor depending on the growth directions and the content of Ge in the SiGe substrate at the $\Gamma$ point in strained Ge. The calculation is performed using an 8-band model, and the crystal has a diamond structure.}
    \label{g}
\end{figure}

In this calculation, we formed an 8-band model and used the parameters listed in Table \ref{para}.
\begin{table}[ht]
\caption{\label{para}The parameters we used in calculation of $g$-factor. Each parameter is defined by experimental.}
    \centering
 \begin{tabular}{lrc} \hline
   parameter & value & Ref. \\ \hline
   $E_c-E_v$ & $0.898$\;\;[eV] & \cite{Semipara} \\
   $\Delta$ & $0.29$\;\;[eV] & \cite{Semipara} \\
   $a_c-a_v$ & $-8.97\pm0.16$\;\;[eV] & \cite{deformation} \\
   $b$ & $-1.88\pm0.12$\;\;[eV] & \cite{deformation} \\
   $C_{11}$ & 12.40\;\;[10$^{11}$ dyn cm$^{-2}$] & \cite{Semipara} \\
   $C_{12}$ & 4.13\;\;[10$^{11}$ dyn cm$^{-2}$] & \cite{Semipara} \\
   $C_{44}$ & 6.83\;\;[10$^{11}$ dyn cm$^{-2}$] & \cite{Semipara} \\ 
   $A$ & -13.38 & \cite{Semipara} \\
   $B$ & -8.57 & \cite{Semipara} \\
   $|C|$ & 12.78 & \cite{Semipara} \\ \hline
\end{tabular}
\end{table}
$A$, $B$ and $|C|$ are valence band parameters used in the $\textbf{k}\cdot\hat{\textbf{p}}$ perturbation method\cite{luttinger1955motion}. We can obtain the momentum matrix elements $p_{kl}^i$ between the conduction band and the valence bands such as HH, LH, and SP, from these parameters using the formula introduced by Dresselhaus, Kip, and Kittel \cite{DKK}. As a result, we obtained effective $g$-factors as shown in Fig.~\ref{g}.
\section{Disscussions and Conclusion}
\begin{figure}[ht]
    \centering
    \includegraphics[width=0.8\linewidth]{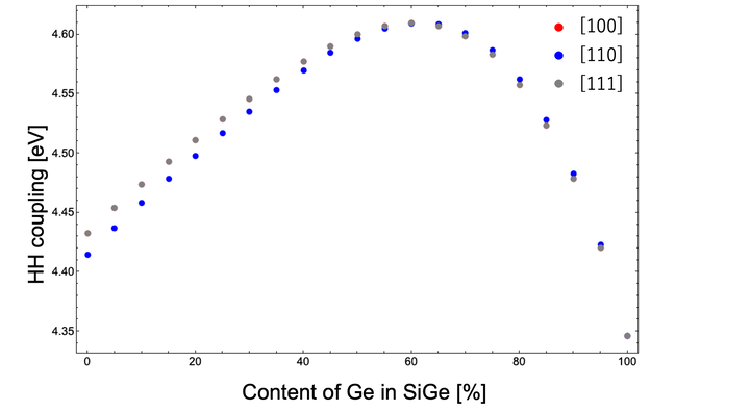}
    \caption{The absolute value of HH coupling $p_{S n}^x p_{nS}^y-p_{S n}^y p_{nS}^x$ depending on the growth directions and the content of Ge in the SiGe substrate at the $\Gamma$ point in strained Ge. The calculation is performed using an 8-band model, and the crystal has a diamond structure. S represents the conduction band and HH represents the valenceband}
    \label{HH}
\end{figure}

We investigated the electron $g$-factor at the $\Gamma$ point in strained Ge layers caused by the SiGe substrate. As shown in Fig.~\ref{g}, the absolute value of the effective $g$-factor decreases with decreasing Ge content in the substrate for all growth directions except for 80 to 100\% of Ge content in the [100] direction. In the region, we observed a slight increase in the absolute value of the electron $g$-factor as the Ge content decreased, with the change being on the order of 0.01. This is because HH coupling increases with decreasing Ge content as shown Fig.\ref{HH}.

Moreover, compared to [110] and [111] directions, the [100] direction exhibits a much more gradual decrease due to the larger perpendicular strain, which minimizes the deformation of the lattice volume. Additionally, we found that the $g$-factor ranges from $-3.0$ to $-1.4$, which is about ten times larger than the hole in-plane $g$-factor ($0.16\sim0.26$) in Ge quantum wells grown on Si$_{0.2}$Ge$_{0.8}$ substrates \cite{Hendrickx2021}. This significant separation makes the system suitable for photon-spin coherent transduction. 

We would like to comment on the factor $D_{111}$, as described in the work by Rideau \etal. \cite{D111} Their expression $D_{111} = (2C_{11} + 4C_{12} - 4C_{44})/(C_{11} + 2C_{12} + 4C_{44})$ is different from ours, suggesting a need for further investigation.

 In all crystal growth directions, there is a sufficient difference between the $g$-factor of conduction band electrons and that of holes, indicating that quantum dots using any crystal growth direction have the potential as quantum memories. Moreover, the $g$-factor difference remains approximately tenfold regardless of the Ge content in the SiGe substrate, allowing for significant flexibility in material design.

In conclusion, we investigated the electron $g$-factor in strained Ge layers on SiGe substrates, analyzing the effects of varying Ge content and growth directions ([100], [110], [111]). The $g$-factor in these orientations ranges from $-3.0$ to $-1.4$.
As a prospect, future work should include the examination of the effects of quantum confinement in the quantum well structure.

\ack
This research was supported by Grant-in-Aid for Scientific Research (S) [23H05458] from JSPS and the JST Moonshot R$\&$D-MILLENNIA Program Grant No. JPMJMS2061\\

 \bibliographystyle{iopart-num}
 \bibliography{iopart-num.bib}
 \nocite{*}

\end{document}